\documentclass[10pt,a4paper,twocolumn,aps,prb,longbilbiography,floatfix,superscriptaddress,showpacs]{revtex4-1}

\usepackage[utf8]{inputenc}
\usepackage[english]{babel}
\usepackage{dirtytalk}
\usepackage{amsfonts}
\usepackage{amsmath}
\usepackage{amssymb}
\usepackage{bbold}
\usepackage[dvipsnames,svgnames,table]{xcolor}
\usepackage{graphicx}
\usepackage{fancyhdr}
\usepackage{float}
\usepackage{braket}
\usepackage{float}
\usepackage[caption=false]{subfig}

\usepackage{hyperref}
\hypersetup{
    pdfstartview={FitH},
    colorlinks=true,    
    linkcolor=NavyBlue, 
    citecolor=Maroon,   
    filecolor=NavyBlue, 
    urlcolor=NavyBlue   
}

\makeatletter
\def\footnoterule{\kern -10pt
    \hrule \@width 100pt \kern 10pt} 
\makeatother

\makeatletter
    \def\bbl@set@language#1{%
      \edef\languagename{%
        \ifnum\escapechar=\expandafter`\string#1\@empty
        \else\string#1\@empty\fi}%
      \@ifundefined{babel@language@alias@\languagename}{}{%
        \edef\languagename{\@nameuse{babel@language@alias@\languagename}}%
      }%
      \select@language{\languagename}%
      \expandafter\ifx\csname date\languagename\endcsname\relax\else
        \if@filesw
          \protected@write\@auxout{}{\string\select@language{\languagename}}%
          \bbl@for\bbl@tempa\BabelContentsFiles{%
            \addtocontents{\bbl@tempa}{\xstring\select@language{\languagename}}}%
          \bbl@usehooks{write}{}%
        \fi
      \fi}
    \newcommand{\DeclareLanguageAlias}[2]{%
      \global\@namedef{babel@language@alias@#1}{#2}%
    }
\makeatother
\DeclareLanguageAlias{en}{english}

\def \newpar{\vspace{10pt}}

\def \cedenna {CEDENNA, Avda. Ecuador 3493, Santiago, Chile.}
\def \fcfm {Departamento de F\'isica, FCFM, Universidad de Chile, Santiago, Chile.}

\def \utsfm {Universidad Técnica Federico Santa María.}

\bibpunct{[}{]}{,}{n}{}{}  

\begin{document}

    \title{Theory of magnetism in the van der Waals magnet $\text{CrI}_3$}
    \author{R. Jaeschke-Ubiergo}
        \affiliation{\fcfm}
        
    \author{E. Su\'arez Morell}
          \affiliation{\utsfm}

\author{A. S. Nunez}
        \affiliation{\fcfm}
        \affiliation{\cedenna}
        
\begin{abstract}
   We study the microscopical origin of anisotropic ferromagnetism in the van der Waals magnet $CrI_3$. We conclude that the nearest neighbours exchange is well described by the Heisenberg-Kitaev-$\Gamma$ (HK$\Gamma$) model, and we also found a nonzero Dzyaloshinskii–Moriya interaction (DMI) on next nearest neighbours. Both Kitaev and DMI are known to generate a non-trivial topology  of the magnons in the honeycomb lattice, and have been used separately to describe the low energy regime of this material. We discuss that including one or the other leads to different signs of the Chern's number. Furthermore, the topological gap at $\boldsymbol{K}$-point seems to be mainly produced by DMI, despite it is one order of magnitude smaller than Kitaev. Finally, we show that by applying an external electric field perpendicular to the crystal plane, it is possible to induce DMI on nearest neighbors, and this could have consequences in non-collinear spin textures, such as domain walls and skyrmions.
\end{abstract}

\maketitle

\section{Introduction}

The two-dimensional wan der Waals magnet $CrI_3$ has attracted interest in the last few years, as well as the rest of the Chromium trihalides \cite{soriano2020magnetic}. In general, two-dimensional crystals with intrinsic magnetism have potential applications in several technologies, such as sensing or data storage \cite{soumyanarayanan2016emergent}. Ferromagnetic semiconductors are relevant in spintronics because these compounds have a sizeable spin-flip length, which is favorable for laterally patterned spin devices \cite{felser2007spintronics}. $CrI_3$ is a ferromagnetic semiconductor with a known gap of $1.2 \text{ eV}$, and a Curie temperature of $68 \text{ K}$ \cite{dillon1965magnetization} (both measured in bulk). 

Few years ago, a monolayer of $CrI_3$ was synthesized, and its magnetic order was demonstrated using magneto-optical Kerr effect microscopy \cite{huang2017layer}. In the single-layer limit, the estimated Curie temperature is 45 K, a bit lower than the bulk material's critical temperature. One of the exciting features of this material is its layer dependent magnetic behavior \cite{huang2017layer,Morell_2019}. Also, it has been used as a ferromagnetic substrate to control spin and valley dynamics\cite{zhong2017van}.

Magnetic anisotropy is necessary to explain the observed out-of-plane ferromagnetic(FM) order in $CrI_3$ \cite{mermin1966absence}; several models have been used with that purpose. The $XXZ$ model was the first proposed \cite{lado2017origin} to explain the FM order. After that, other authors proposed that a single-layer $CrI_3$ is a Heisenberg-Kitaev magnet, with a positive Kitaev constant \cite{xu2018interplay, lee2020fundamental, stavropoulos2019microscopic, pizzochero2020magnetic}. Moreover, recently, analytical calculations suggested a negative Kitaev constant instead. \cite{stavropoulos2020magnetic}. Heisenberg exchange $ J $, Kitaev coupling $ K $, and symmetric off-diagonal exchange $\Gamma$ are compatible with the crystal's symmetries and have also been used to describe magnetism in honeycomb Iridium oxides \cite{rau2014generic, yamaji2014first}. Furthermore, next-nearest neighbors (NNN) Dzyaloshinskii–Moriya interaction (DMI) it is also allowed by symmetry \cite{chen2018topological}, and the component of the DM vector which is collinear to the magnetization played the same role in magnons as the SOC in the Kane-Mele's model \cite{kane2005quantum}

Parallel to the development of electronic and magnetic models, spin excitations on $CrI_3$ has been also studied on the single-layer limit\cite{stavropoulos2020magnetic,da2020topological, chen2018topological, aguilera2020topological}, on bilayers \cite{ortmanns2020magnon} and heterostructures \cite{hidalgo2020magnon}. Most prominent results in this aspect are the anisotropy gap $\Delta_{\Gamma}$ of the lower band at $\boldsymbol{\Gamma}$-point, and the $\boldsymbol{K}$-point gap, which also has been remarked as a signature of a non-trivial topology in the magnons. Among the experimental evidence of magnons, there are inelastic neutron scattering experiments on multilayered samples \cite{chen2018topological}, which estimates the gap at Dirac point as $4 \text{ meV}$, and the $\boldsymbol{\Gamma}$-point gap in the order of $2 \text{ meV}$. In the same work, exchange constants for nearest negihbours and next nearest neighbours are $2.01 \text{  meV}$ and $0.16 \text{ meV}$ respectively. Recently, magnons in single-layer $CrI_3$ were directly observed through  magneto-Raman spectroscopy \cite{cenker2020direct}. The spin wave's gap was estimated to be $\Delta_{\boldsymbol{\Gamma}} \approx 0.3 \text{  meV}$, and the exchange $J\approx 2.83 \text{  meV}$.

In this work, we propose a strategy to deduce the spin Hamiltonian, which governs the magnetic degrees of freedom in single-layer $CrI_3$. Furthermore, through this, we aim to shed some light on understating the magnetic anisotropy on this material.
Our technique's basic idea is to use a Green's functions method \cite{liechtenstein1987local, mazurenko2005weak}, to calculate the energy variation of the ground state when the magnetization on each $Cr$ site is rotated at a small angle on an arbitrary axis. After that, we will map that energy variation into a 3/2-spin Hamiltonian. Similar methods has been used before in the estimation of Heisenberg exchange constants \cite{korotin2015calculation, mazurenko2006electronic}. This technique has been utilized to calculate the orbital-resolved contribution to the exchange in single-layer $CrI_3$ \cite{kashin2019orbitally}. However, we extend the formalism presented in \cite{liechtenstein1987local, mazurenko2005weak}, to include second-order terms in the spin-orbit coupling (SOC). The way we calculate the energy variations makes this technique useful to distinguish between the different couplings proposed to describe the magnetism in $CrI_3$, from an ab initio point of view.

From the magnonic picture, we calculate the spectrum and the Chern's number associated with each band. A non-trivial topology is revealed, as a consequence of the SOC of ligands. We discuss that this non-trivial topology results from an interplay between NN Kitaev coupling and NNN Dzyaloshinskii–Moriya exchange.
Electrical control of magnetic properties of $CrI_3$ has also aroused interest \cite{huang2018electrical, jiang2018controlling}. Although single-layer $CrI_3$ possesses an inversion center, this symmetry can be broken by applying an out-of-plane electric field. This, together with the SOC of the ligands, generates a NN-DMI. As long as there is a planar component of the DM vector, it might be possible to generate skyrmions by applying a voltage difference, as proposed in \cite{behera2019magnetic}. We also calculate this NN antisymmetric exchange as a function of the applied electric field.

\begin{figure}
    \centering
    \includegraphics[scale=0.2]{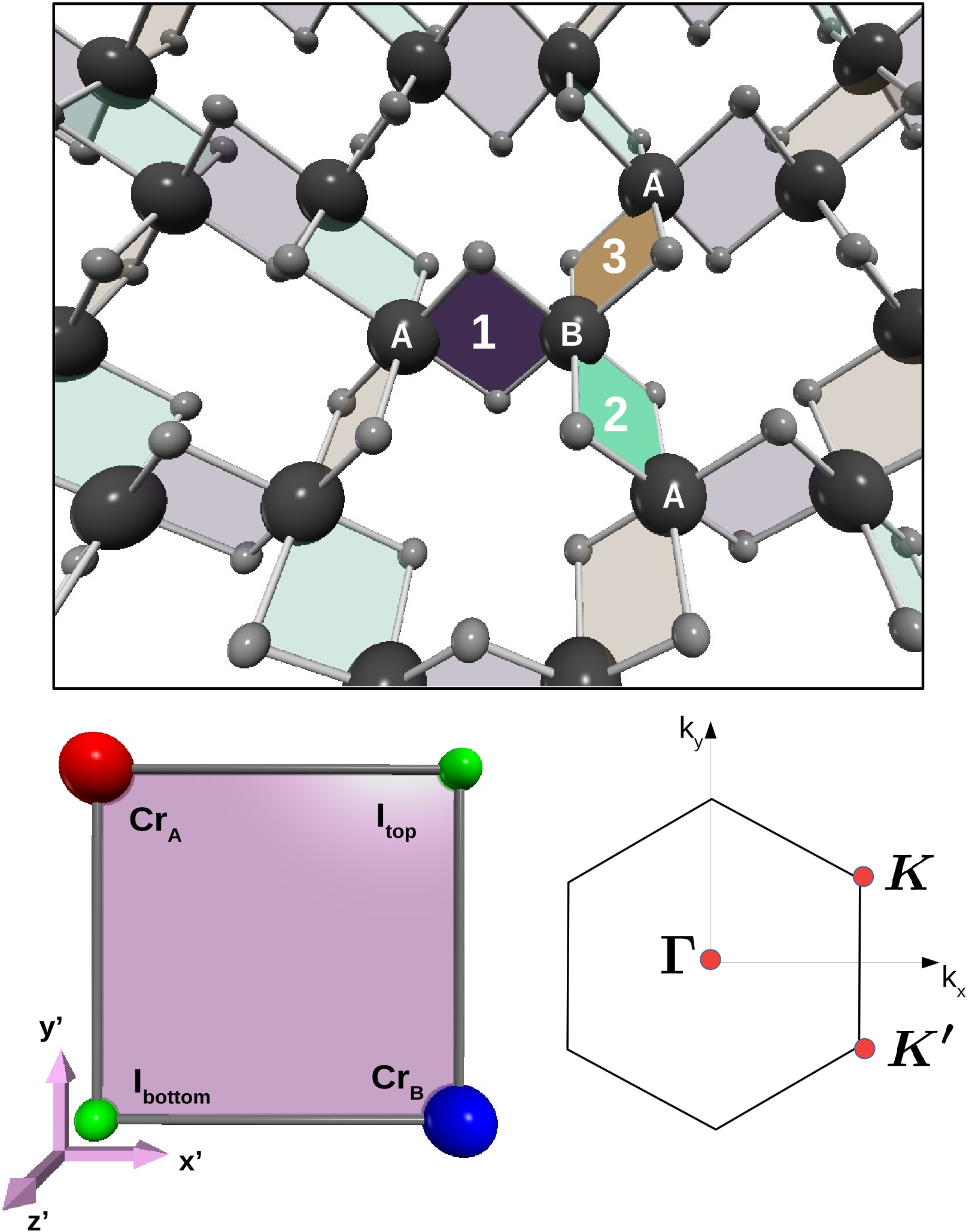}
    \caption{\textbf{(upper panel)} Top view of the $CrI_3$ layer, with different plaquettes $Cr_2I_2$ highlighted in color. Three kind of NN-links are enumerated from 1 to 3. \textbf{(lower-left panel)} Plaquette 1 in the coordinate system $\{x',y',z'\}$. Each plaquette is composed by two $Cr$ sites ($A$ (red) and $B$ (blue)), and two Iodines (green) $I_{top}$ and $I_{bottom}$. Axis $z'$ is normal to the plaquette. Plaquettes 2 and 3 can be obtained by a (111) three fold rotation, or equivalently by permuting the axes $x'$, $y'$, $z'$. \textbf{(lower-right panel)} First Brillouin zone, with symmetry points $\boldsymbol{\Gamma}$ and $\boldsymbol{K}$ highlighted.}
    \label{fig:crystal}
\end{figure}

\section{Methods}

Our approach to obtain the Hamiltonian goes as follows: We performed first a Density Functional theory (DFT) calculation without including the local electronic repulsion, and spin orbit coupling (SOC), from DFT calculations we obtain a tight binding model where we introduce later these interactions as suggested by Kanamori\cite{kanamori1963electron,georges2013strong}, after that using a self-consistent Hartree-Fock approximation we found the minimal energy. Spin orbit coupling is included pertubatively in the deduction of the spin Hamiltonian.

The DFT calculations were done using Quantum Espresso (QE)\cite{Giannozzi2009}, the projector augmented wave (PAW) method\cite{p1_Corso2014} and PBE exchange correlation functional\cite{PBE} were employed.  A grid
of $6 \times 6 \times 1$ k points was used to relax the structures and a finer grid of up to $16 \times 16 \times 1$ to obtain the total energies and band structures with a convergence threshold of $10^{-8}$ eV. This allows us to obtain the band structure of $CrI_3$ when both, local electronic repulsion, and SOC are neglected. 

The resultant band structure is projected into a set of Maximally Localized Wannier Orbitals (MLWO)\cite{Pizzi2020}, which obey the symmetries of p-like orbitals in $I^{-1}$ anions, and d-like orbitals in the case of $Cr^{+3}$ cations (see appendix A). This projection gives us an effective multi-orbital tight-binding model. 

In order to include the repulsion between electrons on the same $Cr$ site, we use the Hubbard-Kanamori model \cite{kanamori1963electron,georges2013strong}, which has been extensively applied to describe magnetism emerging from Hund's rules in transition metal compounds \cite{kanamori1963electron,georges2013strong,stavropoulos2020magnetic,sherman2020hubbard,aron2015analytic,de2011janus}. This model includes the repulsion $U$ between electronic densities on the same orbital, the repulsion of electrons on different orbitals $U'$, and an intra-atomic exchange $J_H$, known as the Hund's coupling. For $d$-like orbitals, assuming spherical symmetry, this couplings are related by $U' = U - 2J_H$.

\begin{figure}
    \centering
    \includegraphics[scale=0.28]{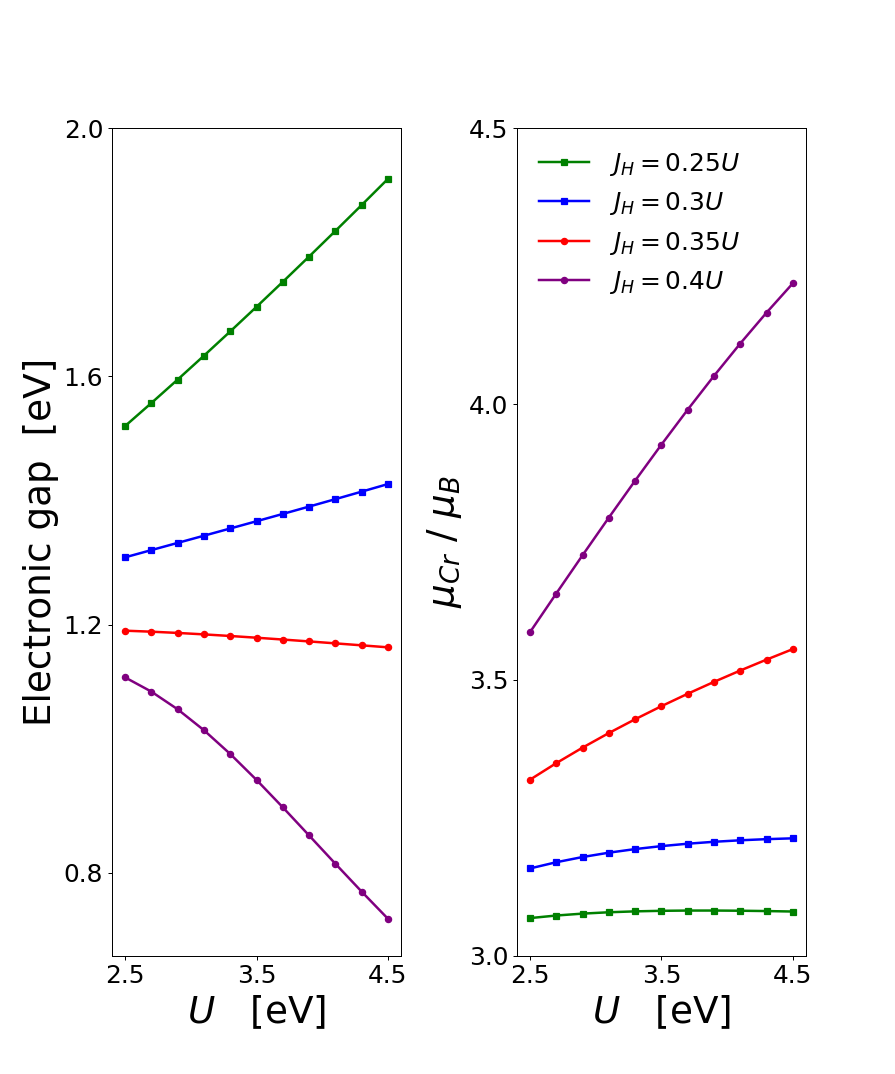}
    \caption{\textbf{(left)} Electronic gap. \textbf{(right)} Magnetic moment on each $Cr$ site. Both quantities are shown as function of the Hubbard parameter $U$, and several ratios $J_H / U$ are included.}
    \label{fig:fermi_gap}
\end{figure}

We treat the Hubbard-Kanamori model self-consistently, using the Hartree-Fock approximation. We obtain a ferromagnetic semiconductor with an electronic gap width $\Delta$ shown in \ref{fig:fermi_gap} as a function of the parameter U. The magnetic moment in the unit cell is 6$\mu_b$ for all studied combination of $U$ and $J_H$. However, the magnetic moment on each $Cr^{+3}$ varies in the range $3.0-4.2$ eV, as it is shown in the right panel of figure \ref{fig:fermi_gap}. $I^{-}$ anions acquire a small magnetization $\mu_I=(\mu_{cr}-3\mu_{B})/3$ in the opposite direction of the spin polarization. Of course, the choice of $U$ and $J_H$ will modify quantitatively the results obtained in the following sections. However, we performed calculations for several values of the Kanamori parameters to show which qualitatively results are robust.

We invoke the magnetic force theorem, as has been previously done \cite{liechtenstein1987local,mazurenko2005weak,mazurenko2006electronic,korotin2015calculation}, to calculate the functional derivative of the ground state energy with respect to the magnetization field. This energy variation can then be mapped into a spin Hamiltonian, which will describe approximately the magnetic degrees of freedom of the single-layer $CrI_3$.

The SOC will be incorporated as a local potential $H^{SO}_{k} = \lambda \mathbf{L}\cdot \mathbf{S}$ on Iodine sites, and we will treat it perturbatively up to second order in $\lambda$. It has been widely argued that the spin-orbit of ligands plays a central role in describing this material \cite{lado2017origin,molina2020magneto}, and for this reason, we neglect SOC in the magnetic chromium atoms. 
 
 The details of the calculation, are provided in the appendix B. The energy variation can be written as:
 \begin{equation}
     \Delta E =  -\sum_i \delta \mathbf{S}_i \cdot \mathcal{A'}_{ii} \cdot \delta \mathbf{S}_i -\frac{1}{2}\sum_{i < j} \delta \mathbf{S}_i\cdot \mathcal{J}_{ij} \cdot \delta \mathbf{S}_j \text{  ,}
     \label{DeltaE}
 \end{equation}
where the first term is an onsite parameter, which includes contributions that could come from a magneto-crystalline anisotropy or a Weiss field associated with the exchange. The second term is an inter-site contribution coming from exchange. Note that we are using $\delta \mathbf{S}_i=(S_i^x, S_i^y)$ as the variation of the magnetization field with respect to the ground state, and for this reason, only the $x$ and $y$-components of the exchange matrix $\mathcal{J}_{ij}$ are involved in \eqref{DeltaE}.

\section{Results and discussion}
\subsection{Spin Hamiltonian}
The previous section explained our methodology to obtain an effective tight-binding Hamiltonian, including the local electronic repulsion. 
We will assume that the ground state's magnetization is polarized along the $\mathbf{\hat{z}}$ axis, perpendicular to the crystal plane. Then, we calculate the energy variation when the magnetization at each $Cr^{+3}$ site is rotated arbitrarily by a small angle of $\delta \phi_i$.
 It is essential to understand that the above methodology allows us to calculate the $x$ and $y$ components of the exchange matrix $\mathcal{J}_{ij}$ and the anisotropy matrix $\mathcal{A}_{ii}$. To illustrate this fact, let us consider the following general spin Hamiltonian:
\begin{equation}
    H_S = -\frac{1}{2}\sum_{\substack{i,j \\(i\neq j)}} \mathbf{S}_i\cdot \mathcal{J}_{ij} \cdot \mathbf{S}_j - \sum_i \mathbf{S}_i \cdot \mathcal{A}_{ii}\cdot \mathbf{S}_i\text{ ,}
    \label{Hs}
\end{equation}

where $\mathbf{S}_i$ is the magnetic moment at the $Cr$ site $i$. The first term considers a generalized exchange interaction. If $\mathcal{J}_{ij}$ were proportional to the identity, it would describe an isotropic Heisenberg exchange, if $\mathcal{J}_{ij}$ had an anti-symmetric part, it would be a Dzyaloshinskii-Moriya interaction (DMI) term and any other symmetric nonzero element in $\mathcal{J}_{ij}$ would represent exotic anisotropies in the exchange process. Similarly, matrix $\mathcal{A}_{ii}$ represents a generalized magneto-crystalline anisotropy. In our case with a $\mathbf{\hat{z}}$-easy axis, the only nonzero element of the onsite matrix should be $\mathcal{A}
^{zz}_{ii}$.

The formalism discussed in appendices B and C, allows us to obtain the $zz$ component of onsite matrix $\mathcal{A}_{ii}$, and the transverse components of the exchange matrices $\mathcal{J}_{ij}$ in \eqref{Hs}. However, if we want to obtain all components, we can repeat the calculation, but changing the polarization of the magnetization. If we chose other polarization axis, let us say $\mathbf{\hat{z}}'$, we could apply all the formalism up to this point, to calculate the $2\times2$ perpendicular block $x', y'$ (relative to $z'$) of each exchange matrix.

The Hubbard-Kanamori model displays full rotational symmetry in the spin space, and the single-electron Hamiltonian obtained by the DFT calculation is spin-diagonal.  When we solve the self-consistent equation, in the Hartree-Fock approximation, we choose a given polarization axis ($\mathbf{\hat{z}}$ if we are looking for the ground state), and this choosing breaks the rotational symmetry of the electronic Hamiltonian. However, prior to this choice, the Hamiltonian has rotational symmetry in its spin components. It is only when we include the SOC, that this symmetry is broken. Because of this, to globally rotate the magnetization, we just have to rotate the spin-orbit contribution $H^{SO}\rightarrow R^{\dagger}H^{SO}R$ (defined in appendix B), with $R$ being the required $SU(2)$ rotation.

The exchange coupling between NN $Cr^{+3}$ is dominated by superexchange paths through $I^{-}$ ligands \cite{xu2018interplay}. Therefore, the natural coordinate systems to calculate the exchange matrices are the ones aligned with the $Cr_2I_2$ plaquettes. The three different kinds of plaquettes are shown with different colours in figure \ref{fig:crystal}. The coordinate system $\{x', y', z' \}$ is aligned with plaquette 1. $x'$ and $y'$ axes are pointing on the direction of $Cr-I$ links, and $z'$ axis is normal to the plaquette. This coordinate system can be used to compute the $\{x', y'\}$ components of $\mathcal{J}_{1}$, $\mathcal{J}_2$ and $\mathcal{J}_3$. Then, a (111)-three-fold rotation allows us to permute the axes, and polarize the magnetization on $\mathbf{\hat{x}'}$. In this rotated system we calculate $\{y', z' \}$ components of all exchange matrices. Finally, we apply another rotation to choose $\mathbf{\hat{y}}$ as the polarization axis, and we obtain the $\{ z', x'\}$ components. This methodology serves us to calculate completely each exchange matrix. Finally, we express $\mathcal{J}_1$ in the basis $\{x',y',z'\}$: (see fig. \ref{fig:crystal})

\begin{equation}
    \mathcal{J}_1 = \left(\begin{aligned}
        J \;\; && \Gamma_{xy} && \Gamma_z \;\; \\
        \Gamma_{xy} && J\; && \Gamma_z \;\; \\
        \Gamma_z && \Gamma_z && J+K
    \end{aligned} \right)\text{  .}
    \label{J1}
\end{equation}

The exchange matrices for the other two links can be obtained by a (111) three fold rotation on the $\{x', y', z' \}$ coordinate system, or equivalently, by a three fold rotation around the $\mathbf{\hat{z}}$ axis, normal to the crystal plane. We recognize $J$ in the diagonal, as the usual Heisenberg exchange. However, the third spin-component ($z'$ in this case), interact with a different exchange constant, which is materialized through the Kitaev's coupling $K$. If the super-exchange paths dominate the exchange processes, it is natural to think that the local symmetry of the $Cr_2 I_2$ plaquette influences it. Moreover, on plaquette 1, $x'$ and $y'$ components are identical, but there is no reason to think that the normal $z'$ component will behave such as the parallel components. In fact, when the SOC is included on Iodine atoms, the lack of spatial rotational symmetry on the plaquette is transferred to the spin degrees of freedom, and a Kitaev's coupling appears on each NN-link. 

We can understand the couplings $\Gamma_z$ and $\Gamma_{xy}$ with a similar reasoning . In figure \ref{fig:crystal}, we see that each $Cr$ atom is surrounded by 6 Iodine atoms, forming an octahedral environment. If this octahedron were perfect, the constants $\Gamma_z$ and $\Gamma_{xy}$ would be zero. We verify this behavior, expressing the hoppings obtained from the wannierization procedure, in terms of Slater-Koster parameters \cite{slater1954simplified}. This allows us to move the centers of the atomic sites simulating a tight binding Hamiltonian for a perfect octahedral environment. We found then that the exchange matrices, in the basis $\{x',y',z'\}$ were all diagonal. The slight deformation of this octhaedron in the actual material, is the reason behind constants $\Gamma_z$ and $\Gamma_{xy}$. This is in agreement with a recent calculation that parametrizes the deformation of the octahedral environment \cite{stavropoulos2020magnetic}. This deviation from the NN cubic symmetry on $CrI_3$ has been reported and incorporated in several other works \cite{lado2017origin, zhang2019super,gu2019observation}. 

The couplings $K$ and $\Gamma$, which represent anisotropies in the exchange interaction, are contained inside the constants $\alpha_{ij}$ and $\beta_{ij}$, we define them in the appendix B. This terms only arise when second order process in the SOC are included. Then, for small spin orbit coupling, the Kitaev's constant  depends on $\lambda$ as $K \sim \lambda^{2}$, and the same for $\Gamma_z$ and $\Gamma_{xy}$, as was proposed in reference \onlinecite{xu2018interplay}.

We found also a non-negligible Heisenberg exchange between next nearest neighbours (NNN) and an anti-symmetric exchange between NNN. Finally, the spin Hamiltonian, which describe the low energy regime in the single-layer $CrI_3$ is:
\begin{equation}
    H = H_{HK\Gamma} +H_{nnn} - A_0 \sum_i (S_i^z)^2 \text{  ,}
    \label{HS_CrI3}
\end{equation}
where $H_{HK\Gamma}$ is the nearest neighbour Hamiltonian, defined by:
\begin{equation}
    H_{HK\Gamma} = -\sum_{\braket{i,j}}\mathbf{S}_i\cdot \mathcal{J}_{ij}\cdot\mathbf{S}_j
\end{equation}
 and $H_{nnn}$ includes both symmetric and antisymmetric exchanges on NNN:
\begin{equation}
    H_{nnn} = -  \sum_{\braket{\braket{i,j}}}\left( J_{nnn}\mathbf{S}_i \cdot \mathbf{S}_j + \mathbf{d}_{ij}\cdot (\mathbf{S}_i \times \mathbf{S}_j)\right)\text{  .}
\end{equation}
Here,  $\mathcal{J}_{ij}\in\{\mathcal{J}_{1},\mathcal{J}_{2},\mathcal{J}_{3} \}$ represent the Heisenberg-Kitaev-$\Gamma$ model described in \eqref{J1}. The second contribution includes isotropic Heisenberg exchange $J_{nnn}$, and DMI between NNN. The DM vector is defined by $\mathbf{d}_{ij} = \tau_{ij}\left(d_{nnn}^{xy} \mathbf{\hat{l}}_{ij} + \nu_{ij}d_{nnn}^z \mathbf{\hat{z}}\right)$, where $\mathbf{\hat{l}}_{ij}$ is the unitary vector from site $i$ to site $j$, $\tau_{ij}$ is $+1 (-1)$ if $ij$ is a $AA$($BB$)-link. Factor $\nu_{ij}$ also takes values $\pm1$, and alternate its sign on the 6 different NNN links of a given site. Finally, the last term is the contribution of the magneto-crystalline anisotropy. Similar models has been proposed previously \cite{xu2018interplay,lee2020fundamental,stavropoulos2019microscopic}, both by ab initio calculations and by experimental methods.
\begin{figure}[H]
    \centering
    \includegraphics[scale=0.28]{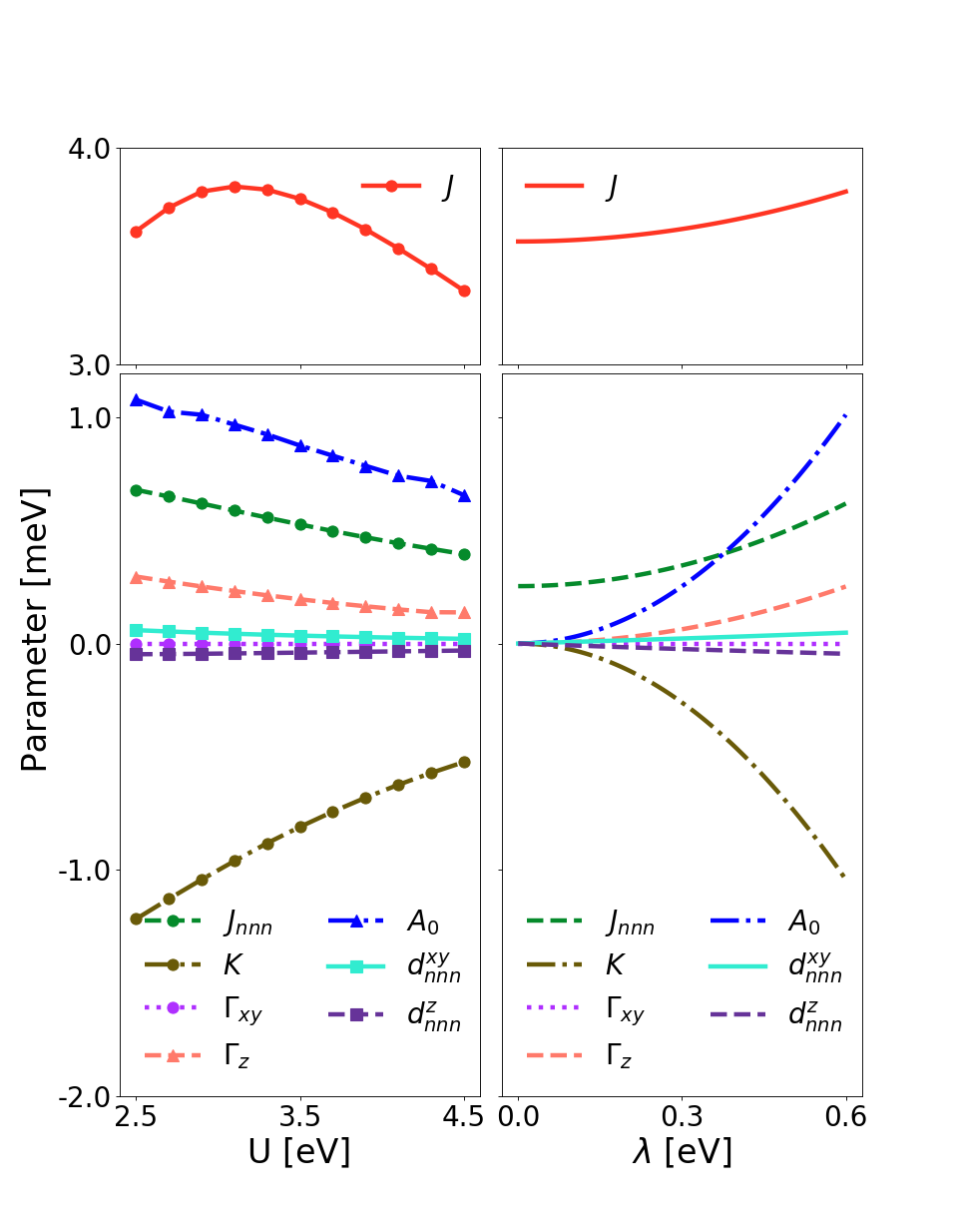}
    \caption{(\textbf{left}) Magnetic constants for different values of the Hubbard parameter $U$, using a fixed value of $\lambda = 0.6 \text{ eV}$. (\textbf{right}) Magnetic constants as functions of the SOC parameter $\lambda$, using $U=2.9 \text{ eV}$. In both plots the hund's coupling is fixed in $J_H=0.25\; U$.}
    \label{fig:mag_coups}
\end{figure}
All magnetic constants involved in the model \eqref{HS_CrI3}, are plotted in figure \ref{fig:mag_coups}, as a function of the Hubbard parameter $U$, and the SOC parameter $\lambda$. For the chosen parameters of $U$, $J_H$ and $\lambda$, the Heisenberg exchange between NN is the dominant coupling in the range $3 - 4 \text{  meV}$. This is consistent with previously reported values \cite{cenker2020direct, chen2018topological}. Furthermore, the NNN Heisenberg exchange is one order of magnitude smaller as in reference  \onlinecite{chen2018topological}. 

We remark the negative sign of the Kitaev constant $K$, in contradiction with the positive value reported in previous articles \cite{xu2018interplay,lee2020fundamental}. However, a similar negative Kitaev coupling was recently found in Ref. \onlinecite{stavropoulos2020magnetic}, where the authors treated the electronic repulsion by means of the Hubbard-Kanamori model.
There is also a nonzero DM interaction on NNN, and it is one order of magnitude smaller than Kitaev interaction, but it can't be neglected because, as we will show in the next section, it significantly contributes to the gap opening at $\mathbf{K}$-point.

\subsection{Topological magnons}
\begin{figure}[H]
    \centering
    \includegraphics[scale=0.37]{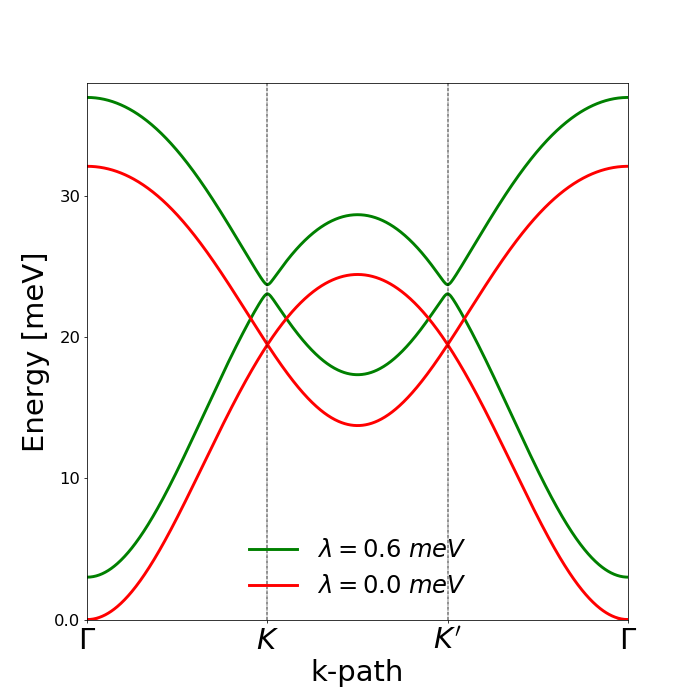}
    \caption{Magnon's bands for $\lambda=0.6 \text{ eV}$. $U=2.9 \text{ eV}$, and $J_H= 0.25 \; U$.}
    \label{fig:magnonic_spectrum}
\end{figure}

\begin{figure}[h]
    \centering
    \includegraphics[scale=0.34]{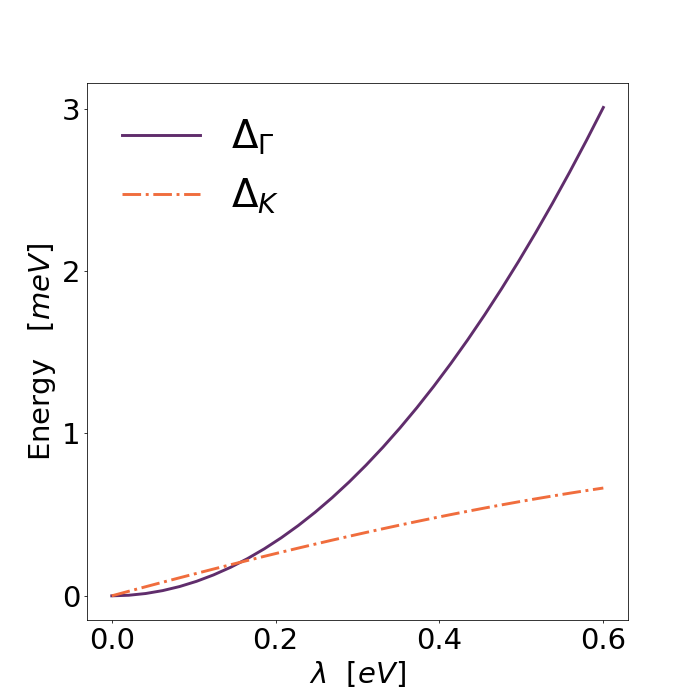}
    \caption{Gaps in the magnonic spectrum at $\boldsymbol{\Gamma}$ and $\boldsymbol{K}$ points as function of the Iodine's SOC parameter $\lambda$}
    \label{fig:magnon_gaps}
\end{figure}

Now, we study the spectrum of the spin excitations around the ground state of the spin model we found in the last section \eqref{HS_CrI3}. We perform a linearized Holstein Primakoff's transformation, and then we diagonalize the quadratic bosonic Hamiltonian using the Colpa's algorithm \cite{colpa1978diagonalization}. In fig. \ref{fig:magnonic_spectrum}, we plotted the magnonic spectrum with and without SOC. Two important features appears when SOC is turned on: (i) The system develops a gap at $\boldsymbol{\Gamma}$-point; (ii) The $\boldsymbol{K}$-point degeneracy is lifted out. These gaps are shown in fig. \ref{fig:magnon_gaps}, as a function of the SOC. The presence of a non-zero Kitaev and NNN-DMI couplings, when SOC is turned-on, added to the gap opening at $\boldsymbol{K}$-point leads the concern to search a non-trivial topological invariant on the magnons. In the present case, this invariant is the integral of the Berry's curvature on the BZ, known as the Chern's number. In fact, several works has already mentioned the topological character of magnons in $CrI_3$ \cite{da2020topological, chen2018topological,aguilera2020topological,hidalgo2020magnon}. Some of these works deal with a Heisenberg-Kitaev model, and other proposes NNN-DMI, whose $\mathbf{\hat{z}}$ component appears in the magnonic Hamiltonian, analogous to the SOC on the Kane-Mele's model\cite{kane2005quantum} in the magnonic Hamiltonian. However, the interplay between Heisenberg-Kitaev-$\Gamma$ and NNN-DMI parts is not clear enough. It is important to calculate the invariant when both terms are present. Furthermore, it could be enlightening to study the Berry's curvature when one of the terms is artificially turned-off. This could give us insight about which mechanism is responsible of the nontrivial topology. 

We calculate the Chern's number on each band according to \cite{shindou2013topological}:

\begin{equation}
    C_j = \frac{i\epsilon_{\mu \nu}}{2\pi} \int_{BZ} d^2 \mathbf{k} Tr[(\mathbf{1}-\mathbf{P}_j)(\partial_{k_{\mu}} \mathbf{P_j})(\partial_{k_{\nu}}\mathbf{P_j})]
    \label{chern}
\end{equation}

Where $\mathbf{P_j}$ is the $4 \times 4$ projector operator of band $j$, defined in \cite{shindou2013topological}. Our calculations indicate that bands plotted in fig. \ref{fig:magnonic_spectrum} has Chern's number $\pm 1$, with the lower band having the positive sign. 

\begin{figure}
    \centering
    \includegraphics[scale=0.22]{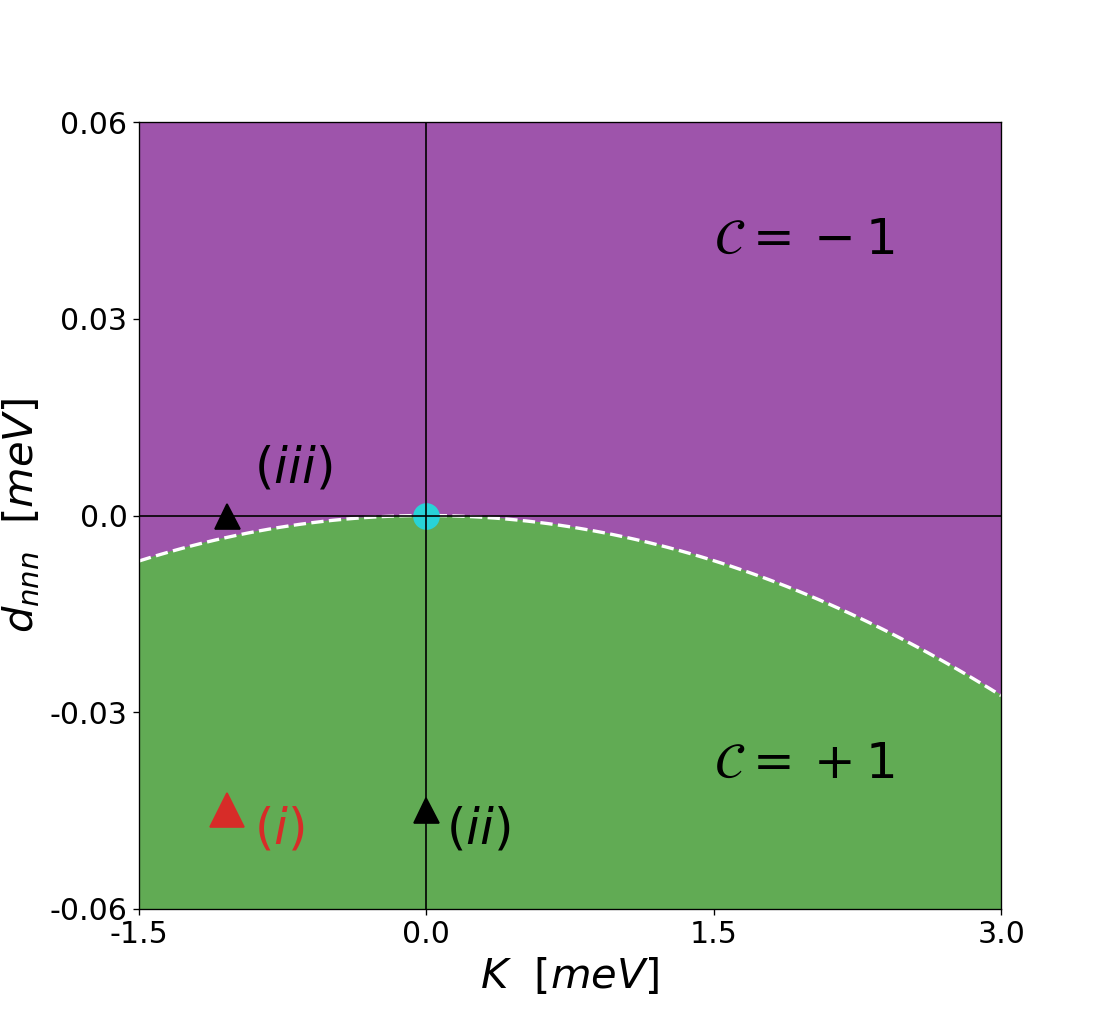}
    \caption{Topological phase diagram with respect to the parameters $K$ and $d_{nnn}^z$. Purple zone has Chern number $-1$, and green zone has Chern number $+1$. The sky blue market at the origin is the only point with $\mathcal{C}=0$. The parameters that we use to describe $CrI_3$ in this article are highlighted with a red triangle (i). Black triangles (ii) and (iii) represent the scenarios with $K=0$ and $d_{nnn}^z=0$ respectively, which are described in the main text.}
    \label{fig:top_diag}
\end{figure}

We can now artificially turn-off some couplings to isolate its effects and analyze the Berry curvature, and the corresponding Chern's number, on three cases: (i) The magnonic model that emerges from the full spin model of equation \eqref{HS_CrI3}; (ii) The same model, but imposing $K=\Gamma_{z}=\Gamma_{xy}=0$; (iii) Same as (ii), but making $d_{nnn}^z=0$ instead. All cases (i) (full model), (ii)(NNN-DMI) and (iii) (Kitaev) presents a nontrivial topology. These 3 scenarios are highlighted in the topological phase diagram of fig. \ref{fig:top_diag}, where we vary the parameters $K$ and $d_{nnn}^z$ keeping the rest of the model intact, we calculate  the Chern's number at each point.

Moreover, in all scenarios the Berry curvature is concentrated around $\boldsymbol{K}$ and $\mathbf{K'}$ valleys, and it is identical in both valleys. It is remarkable the similarities between the berry curvatures of the full model (i) and the NNN-DMI case (ii). It is also surprising that Chern's number in Kitaev case (iii) has the opposite sign with respect to the other two cases. This change of sign is an indicator of a topological phase transition, which is represented by the white line in fig. \ref{fig:top_diag}, and it has important macroscopical consequences. For instance, in a finite sample, the chirality of the edge chiral spin waves depends on the Chern's number sign.

All peaks in the Berry curvature, are indicators of singularities in the parameter space, that appears when a degeneracy (gap closing) occurs. A sharper peak (as in the Kitaev case (iii)), indicates a closer degeneracy, and a smaller gap. This simple reasoning allows us to identify the NNN-DMI as the main responsible for the gap opening. This is quite interesting because, in fact, $d_{nnn}^z$ is one order of magnitude smaller than $K$.

\subsection{Nearest neighbours DMI induced by electric field}
The NN anti-symmetric exchange interaction is identically zero due to the fact that the single-layer $CrI_3$ has inversion symmetry. However, we can break this symmetry by applying an electric field perpendicular to the crystal plane \cite{liu2018electrical,behera2019magnetic}. To model this effect, we introduce an onsite potential $\mp \Delta$ on the Iodine atoms of the top($-$) and bottom $(+)$ planes respectively. 

We perform the self consistent procedure of the Hartree-Fock approximation, using different values of the electric field $E_0$ and calculate $d_{ij}$, which is the anti-symmetric components of the exchange matrix $\mathcal{J}$ (see appendix B). This term can be mapped into a Dzyaloshinskii–Moriya interaction with energy given by:
\begin{equation}
    E_{DMI} = -\sum_{\braket{i,j}} \mathbf{d}_{ij} \cdot (\mathbf{S}_i \times \mathbf{S}_j) \text{  ,}
\end{equation}
where $\mathbf{d}_{ij}=d_{xy}\mathbf{\hat{z}}\times \mathbf{\hat{l}}_{ij} + d_z \mathbf{\hat{z}}$. Here, $\mathbf{\hat{l}}_{ij}$ is the unitary vector from site $i$ to site $j$, and $\mathbf{\hat{z}}$ is the unit vector normal to the crystal periodic plane. Constants $d_{xy}$ and $d_z$ are equal on every NN link, and we found that both have a linear relation with the applied electric field. In such a way that $d_z = c_{z}(\lambda) E_0$ and $d_{xy} = c_{xy}(\lambda) E_0$.

\begin{figure}[H]
    \centering
    \includegraphics[scale=0.4]{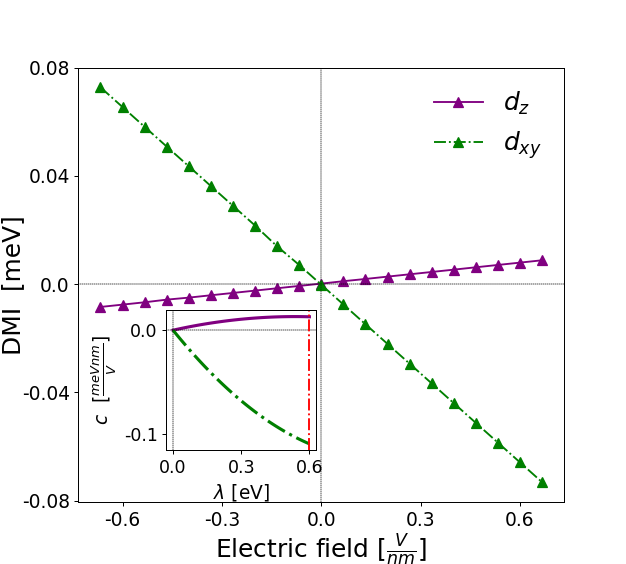}
    \caption{Components of the DM vector, as a function of the electric field, using $\lambda = 0.6 \text{  eV}$. The inset shows the behavior of $c_{xy}(\lambda) = d_{xy} / E_{0}$, and $c_{z}(\lambda) = d_{z} / E_{0}$ as a function of the Iodine's SOC parameter $\lambda$. In this calculation we used $U=2.9$ $eV$ and $J_H / U = 0.25$.}
    \label{fig:dmi}
\end{figure}

 This linear behavior on $E_0$ is expected for reasonable small fields. As can be seen in figure \ref{fig:dmi}, the planar component of the DM vector is greater than the $\mathbf{\hat{z}}$-component. Also, the planar component points in the direction $\mathbf{\hat{z}}\times \mathbf{\hat{l}}_{ij}$, unlike the planar NNN-DMI that points on $\mathbf{\hat{l}}_{ij}$. As only $d_{z}=8.8 \times 10^{-3} \text{ meV}$ contributes to the magnonic spectrum, the change on the latter is almost negligible. However, the planar component $d_{xy}=-7.3 \times 10^{-2} \text{ meV}$ is about 50 times smaller than the NN-exchange, and 10 times smaller than the NNN exchange. It could play an important role in the study of domain walls and magnetic skyrmions. Another mechanism to break the inversion symmetry, and to induce a DMI on $Cr$ atoms, is explored in Janus-monolayers $Cr(I,X)_3$ \cite{xu2020topological}, and $CrGe(Se,Te)_3$ \cite{zhang2020emergence}.

\section{Summary and conclusions}
In this work, we studied the microscopical origin of ferromagnetism in $CrI_3$. To do so, we start from a single electron DFT Hamiltonian, and we incorporate coulomb repulsion using the Hubbard-Kanamori model. Interactions were treated self-consistently with the Hartree-Fock approximation. Once a mean-field Hamiltonian was obtained, we used a Green's functions method to calculate the ground state energy variations when the magnetization is rotated arbitrarily on each site. Spin-orbit coupling was included perturbatively up to second-order as a local potential on Iodine atoms.
We found that a Heisenberg-Kitaev-$\Gamma$ model well describes the magnetic degrees of freedom in the single-layer $CrI_3$ on NN, added to Heisenberg exchange and DM couplings on NNN. The HK$\Gamma$ model on NN is dominated by the exchange $J$, which is found to be $\sim 3.7 \text{meV}$. An important finding is the negative Kitaev coupling, which has also been suggested by analytical calculations \cite{stavropoulos2020magnetic}. The off-diagonal symmetric exchange $\Gamma_z$ appears as a result of the deformation of the octahedral environment of the $I^{-}$ ions surrounding each $Cr^{+3}$ cation. We also found a strong easy axis single-ion anisotropy of $\sim 1 \text{ meV}$. On the other hand, on NNN couplings, the Heisenberg exchange $J_{nnn}$ is one order of magnitude smaller than $J$. There is also a nonzero anti-symmetric exchange $d_{nnn}$, that despite being small, plays an important role in the topological character of spin waves.

We also study the spin excitations of the magnetic ground state, magnons. Our model reproduces the gap at $\boldsymbol{\Gamma}$-point, which prevents the long-range magnetic order from being annihilated by thermal fluctuations. Moreover, we found that Dirac cones degeneracy $\boldsymbol{K}$ and $\boldsymbol{K'}$ is also lifted out. Both gaps are a direct consequence of the SOC of $I^{-}$ anions.

Both Kitaev and NNN-DMI are known to produce nontrivial topologies in the magnonic honeycomb lattice. We study the interplay between them and conclude that NNN-DMI is the main responsible for the gap opening at $\boldsymbol{K}$-point. However, it would be enlightening to study the topological phase diagram in the $K-d_{nnn}$ parameter space theoretically. This could give a better understanding of the sign inversion of the Chern's number when NNN-DMI is added to the HK$\Gamma$ model. As magnetic couplings can be tuned in many ways in two-dimensional Chromium trihalides, a deeper understanding of the topology's origin is needed to design novel applications that takes advantage of dissipationless magnonics \cite{wang2018topological}.

We also showed that an electric field applied perpendicular to the crystal plane breaks the inversion symmetry, this together with the SOC in Iodine atoms, leads to a non-zero nearest neighbor DM constant. The electric field could control non-collinear spin textures, such as skyrmions \cite{behera2019magnetic} and domain walls.

The technique used here proved to be useful to describe magnetic couplings in two-dimensional materials and their hetero-structures. In particular, it is straight forward to apply in other chromium trihalides. We hope that our work also serves as a guide for understanding and controlling other Kitaev materials, such as $\alpha-RuCl_3$, which has been proposed as a spin-liquid candidate \cite{banerjee2017neutron}.

\newpar
\textit{Acknowledgments} - A.S.N. thanks Joaqu\'in Fern\'andez-Rossier for helpful comments. The authors thanks Fondecyt Regular 1190324, R.J.-U. thank to  Financiamiento Basal  para  Centros  Cient\'ificos  y  Tecnol\'ogicos  de  Excelencia FB 0807. Powered@NLHPC: This research/thesis was partially supported by the supercomputing infrastructure of the NLHPC (ECM-02).

\nocite{*}

\onecolumngrid

\counterwithin{figure}{section}
\appendix
\section{Maximally Localized Wannier Orbitals}
The electronic configurations of ions $I^{-1}$ and $Cr^{+3}$ are given by:
\begin{equation*}
    \begin{aligned}
    Cr^{+3} :&& [Ar] 3d^3\text{  ,} \quad\quad\quad\;\\
    I^{-1}  :&&   [Kr]4d^{10}5s^2 5p^6 \text{  .}
    \end{aligned}
\end{equation*}

We are interested in describing the low energy electronic structure of $CrI_3$ monolayer. It is reasonable then to use a model which includes only the outer shell electrons, and then,  project the bands structure only over the orbitals in which these outer electrons are mostly localized. We will use the $5p$ orbitals in $I$ sites: $\ket{p_x}$; $\ket{p_y}$; $\ket{p_z}$. In the case of $Cr$ atoms, the $d$-layer is semi-occupied but for completeness, we will include all $3d$ orbitals: $\ket{d_{xy}}$; $\ket{d_{yz}}$; $\ket{d_{zx}}$; $\ket{d_{x^2-y^2}}$; $\ket{d_z^2}$.

We force the maximally localized Wannier functions representing the selected orbitals to have the angular symmetries of the cubic harmonics listed above. This can be done automatically  by the software Wannier90 \cite{souza2001maximally}. We did not impose any constraint to the radial part of the Wannier functions.

If we include the spin freedom degree, we have 56 states (28 Wannier functions on each unit cell). We selected from ab initio calculations, the states closer to the Fermi level.

\section{Local force theorem and magnetic couplings terms of Green's functions}
 Up to first order in the charge and spin densities, the variation of the ground state energy takes the form:
\begin{equation}
\begin{aligned}
    \Delta E = -\frac{1}{\pi}Im \int_{-\infty}^{E_F}d\epsilon  & Tr\{\delta H G(\epsilon)+ \delta^2 H G(\epsilon) \\
    +& \delta H G(\epsilon)\delta H G(\epsilon)\} \text{  .}
\end{aligned}
    \label{local_force}
\end{equation}

Here, $G(\epsilon) = (\epsilon + i0^{+}- H)^{-1}$ is the Green's function of the collinear system (when SOC is neglected), $\delta H$ and $\delta^{2} H$ are the first and second order variations of the Hamiltonian respectively, when the magnetization at each site $i$ is rotated an arbitrarily small angle $\delta \vec{\phi}_i = (\delta \phi_i^x, \delta \phi_i^y, 0)$. Note that we start from the collinear configuration, with magnetization pointing on $\mathbf{\hat{z}}$-axis, and then, a $z$-rotation does not change the energy. The variations of the Hamiltonian take the form:

\begin{equation}
        \delta H_{ii} = -\frac{\Delta_i}{2}\delta \phi^x_i \sigma_y + \frac{\Delta_i}{2}\phi^y_i \sigma_x \text{  ,}
    \label{deltaH}
    \end{equation}
and 
    \begin{equation}
        \delta^2 H_{ii}  =-\frac{\Delta_i}{4}|\delta \vec{\phi}_i|^2\sigma_z\text{  .}
    \label{delta2H}
\end{equation}

where $\Delta_i = H_{ii}^{\uparrow\uparrow} - H_{ii}^{\downarrow\downarrow}$ is the onsite potential on magnetic $Cr$ sites. Label $i$ represents a given $Cr$ site, so $\Delta_i$ is a $5\times 5$ matrix, whose entries are its projections in the basis of $d$-orbitals (see appendix A).

As discussed in the main text, the spin orbit coupling is incorporated perturbatively up to second order in the SOC parameter $\lambda$. For simplicity, we only consider those terms which are local, and we neglect any inter-site hopping that emerges from the SOC. In this way, by projecting the SOC into the Iodine's $p$-orbitals, we get:

 \begin{equation}
     H_k^{SO} = \frac{\lambda}{2} \left(\begin{aligned}
         0 && -i \sigma_z && i\sigma_y\\
         i \sigma_z && 0 && -i\sigma_x\\
         -i \sigma_y && i\sigma_x && 0
     \end{aligned} \right) \text{  .}
 \end{equation}
 
 The above matrix is represented in the basis $\{ \ket{p_x}, \ket{p_y}, \ket{p_z}\}$, and each entry is a Pauli matrix, accounting the spin components of the  SOC. After the substitution $G \rightarrow G + G^{(1)}+ G^{(2)}$, with $G^{(1)}= GH^{SO}G$, and $G^{(2)}= GH^{SO}GH^{SO}G$, on eq. \eqref{local_force}, the variation in energy can be grouped according to the order in $\delta \phi_i$:
\begin{equation}
    \Delta E = \delta E + \delta^2 E\text{  .}
\end{equation}
Furthermore, each term in the above equation can be grouped again, according to its order in the SOC parameter $\lambda$:

\begin{equation}
    \delta E = \delta E_0 + \delta E_{\lambda} + \delta E _{\lambda^2} \text{  ,}
\end{equation}
\begin{equation}
    \delta^2 E = \delta^2 E_0 + \delta^2 E_{\lambda} + \delta^2 E_{\lambda^2} \text{  .}
\end{equation}
These six different energy contributions are listed below:

\begin{equation}
    \delta E_0 = \frac{-1}{\pi}\int_{-\infty}^{E_F}d \epsilon Tr\{ \delta H G\} \text{  ,}
\end{equation}

\begin{equation}
    \delta E_{\lambda} = \frac{-1}{\pi}\int_{-\infty}^{E_F}d \epsilon Tr\{\delta H G^{(1)} \}\text{  ,}
\end{equation}

\begin{equation}
    \delta E_{\lambda^2} = \frac{-1}{\pi}\int_{-\infty}^{E_F}d \epsilon Tr\{ \delta H G^{(2)}\}\text{  ,}
\end{equation}

\begin{equation}
    \delta^2 E_0 = \frac{-1}{\pi}\int_{-\infty}^{E_F}d \epsilon Tr\{\delta^2 H G+ \delta H G \delta H G  \}\text{  ,}
\end{equation}

\begin{equation}
    \delta^2 E_{\lambda} = \frac{-1}{\pi}\int_{-\infty}^{E_F}d \epsilon Tr\{  \delta^2 H G^{(1)}+\delta H G \delta H G^{(1)}+\delta H G^{(1)}\delta H G\}\text{  ,}
\end{equation}

\begin{equation}
    \delta^2 E_{\lambda^2} = \frac{-1}{\pi}\int_{-\infty}^{E_F}d \epsilon Tr\{ \delta^2 H G^{(2)}+ \delta H G \delta H G^{(2)} + \delta H G^{(1)} \delta H G^{(1)} + \delta H G^{(2)}\delta H G\} \text{  .}
\end{equation}

It is straightforward to verify that $\delta E_0=0$ because $\delta H$ and $G$ are spin-anti-diagonal and spin-diagonal, respectively, so the product has null trace. The remaining terms, will be treated as follows. First, we expand the trace in the site and spin basis. We use $Tr_{L}$ to denote the orbital trace. After replacing the explicit form of $\delta H$ and $\delta^2 H$ of eqs. \eqref{deltaH} and \eqref{delta2H}, we group each term according the spin components involved. For example, $\delta E_{\lambda}$ takes the form:

\begin{equation}
\begin{aligned}
    \delta E_{\lambda} &= -\sum_i h_i^{x(1)} S_i^x + h_i^{y(1)} S_i^y\\
    &= -\sum_i \mathbf{h}_i^{(1)} \cdot \delta \mathbf{S}_i \text{  ,}
\end{aligned}
\end{equation}
when we have defined the components of vector $\mathbf{h}_i^{(1)}$ as:
\begin{equation}
    \begin{aligned}
        h_i^{x(1)} = \frac{1}{2\pi S} Im \int_{-\infty}^{E_F} d \epsilon Tr_l\{\Delta_i(G_{ii}^{(1)\uparrow \downarrow} +G_{ii}^{(1)\downarrow \uparrow} ) \}\text{  ,}\\
        h_i^{y(1)} = \frac{1}{2\pi S} Re \int_{-\infty}^{E_F} d \epsilon Tr_l\{\Delta_i(G_{ii}^{(1)\uparrow \downarrow} -G_{ii}^{(1)\downarrow \uparrow} ) \}\text{  .}
    \end{aligned}
    \label{h1}
\end{equation}\\

In analogous way, $\delta E_{\lambda^2}$ takes the form:
\begin{equation}
    \delta E_{\lambda^2} = -\sum_i \mathbf{h}_i^{(2)}\cdot \delta \mathbf{S} \text{ ,}
\end{equation}
with
\begin{equation}
    \begin{aligned}
        h_i^{x(2)} &= \frac{1}{2\pi S} Im \int_{-\infty}^{E_F} d\epsilon Tr_L \{\Delta_i(G_{ii}^{(2)\uparrow\downarrow}+G_{ii}^{(2)\downarrow\uparrow}) \}\text{  ,}\\
        h_i^{y(2)} &= \frac{1}{2\pi S} Re \int_{-\infty}^{E_F} d\epsilon Tr_L \{\Delta_i(G_{ii}^{(2)\uparrow\downarrow}-G_{ii}^{(2)\downarrow\uparrow}) \} \text{  .}
    \end{aligned}
    \label{h2}
\end{equation}
A term identical to $\delta E$ was obtained in \cite{mazurenko2005weak}, in the context of canted antiferromagnetism.

Now, we explore the second order in spin, and $\lambda$-independent term $\delta^2 E_0$. After some manipulation, it can be written as:

\begin{equation}
    \delta^2 E_0 = \frac{1}{4}\sum_{ij}J_{ij}^{0} |\delta \mathbf{S}_i-\delta \mathbf{S}_j|^2 \text{  .}
\end{equation}

The above expression  has the form of an isotropic Heisenberg exchange, with the exchange constant given by

\begin{equation}
J_{ij}^{0} = \frac{1}{\pi S^2}Im \int_{-\infty}^{E_F} d\epsilon Tr_L \{\Delta_i G_{ij}^{\downarrow}\Delta_j G_{ji}^{\uparrow} \}\text{  .}
\end{equation}

This term coincides with that obtained in \cite{mazurenko2005weak}, except for a factor 2, which appears because the convention of the exchange used in this article. The remaining terms, can be classified in two groups:

\begin{itemize}
 \item[(i)]$ E^{(n)}_{onsite} =\frac{-1}{\pi}Im \int_{-\infty}^{E_F}d\epsilon Tr\{\delta^2 H G^{(n)} \}$ \text{  .}
 \item[(ii)]$E^{(n1,n2)}_{intersite} = \frac{-1}{\pi}Im \int_{-\infty}^{E_F}d\epsilon Tr\{\delta H G^{(n1)}\delta H G^{(n2)} \}$ \text{ .}
\end{itemize}

Here, $n_1$ and $n_2$ reminds the orden in $\lambda$. Onsite contributions are those in which the variation of the Hamiltonian appears only one time, and so, after expanding the trace in the site basis, we obtain onsite terms in the spin Hamiltonian. Explicitly we have

\begin{equation}
    E^{(n)}_{onsite}= \sum_i K_i^{(n)} |\delta \mathbf{S}_i|^2 \text{  ,}
\end{equation}

were we have defined the onsite constant $K_i^{(n)}$ as
\begin{equation}
    K_i^{(n)} = \frac{1}{2\pi S^2} Im\int_{-\infty}^{E_F} d\epsilon Tr_L\{ \Delta_i(G_{ii}^{(n)\uparrow\uparrow}-G_{ii}^{(n)\downarrow\downarrow})\}\text{  .}
\end{equation}

The intersite terms, are those in which the variation of the Hamiltonian appears two times. They are slightly less direct to obtain than the previous ones, but after grouping different components of the exchange processes, we obtain:

 \begin{equation}
     E^{(n_1, n_2)}_{intersite} = \frac{-1}{2} \sum_{ij} \delta \mathbf{S}_i\cdot\left(\begin{aligned}
         J_{ij}^{(n_1,n_2)} + \alpha_{ij}^{(n_1,n_2)} && \beta_{ij}^{(n_1,n_2)} + d_{ij}^{(n_1,n_2)}\\
         \beta_{ij}^{(n_1,n_2)}-d_{ij}^{(n_1,n_2)} && J_{ij}^{(n_1,n_2)} - \alpha_{ij}^{(n_1,n_2)}
     \end{aligned} \right)\cdot \delta \mathbf{S}_j\text{  .}
     \label{intersite}
 \end{equation}

The above matrix is a $2\times 2$ block of the full exchange matrix. $J_{ij}^{(n_1, n_2)}$ is the isotropic part of the exchange. The term $d_{ij}^{(n_1, n_2)}$ opens the possibility to obtain an anti-symmetric exchange (DMI), and finally, $\alpha_{ij}^{(n_1, n_2)}$ and $\beta_{ij}^{(n_1, n_2)}$ are more exotic terms, which represent anisotropies in the exchange procesess. All these magnetic constants are defined, in terms of the Green's functions, as follows:

 \begin{equation}
 \begin{aligned}
     J_{ij}^{(n_1,n_2)} &= \frac{1}{2\pi S^2} Im \int_{-\infty}^{E_F}d\epsilon Tr_L\{\Delta_i G_{ij}^{(n1)\downarrow\downarrow}\Delta_j G_{ji}^{(n_2)\uparrow\uparrow} + \Delta_i G_{ij}^{(n_1)\uparrow\uparrow}\Delta_j G_{ji}^{(n_2)\downarrow\downarrow} \}\text{  ,}\\
      d_{ij}^{(n_1,n_2)} &= \frac{1}{2\pi S^2} Re \int_{-\infty}^{E_F}d\epsilon Tr_L\{\Delta_i G_{ij}^{(n1)\downarrow\downarrow}\Delta_j G_{ji}^{(n_2)\uparrow\uparrow} - \Delta_i G_{ij}^{(n_1)\uparrow\uparrow}\Delta_j G_{ji}^{(n_2)\downarrow\downarrow} \}\text{  ,}\\
    \alpha_{ij}^{(n_1,n_2)} &= \frac{1}{2\pi S^2} Im \int_{-\infty}^{E_F}d\epsilon Tr_L\{\Delta_i G_{ij}^{(n_1)\uparrow\downarrow}\Delta_jG_{ji}^{(n_2)\uparrow\downarrow} + \Delta_iG_{ij}^{(n_1)\downarrow\uparrow}\Delta_jG_{ji}^{(n2)\downarrow\uparrow} \}\text{  ,}\\
      \beta_{ij}^{(n_1,n_2)} &= \frac{1}{2\pi S^2} Re \int_{-\infty}^{E_F}d\epsilon Tr_L\{\Delta_i G_{ij}^{(n_1)\uparrow\downarrow}\Delta_jG_{ji}^{(n_2)\uparrow\downarrow} - \Delta_iG_{ij}^{(n_1)\downarrow\uparrow}\Delta_jG_{ji}^{(n2)\downarrow\uparrow} \}\text{  .}\\
    \end{aligned}
 \end{equation}\\

Now, we are ready to calculate $\delta^2 E_{\lambda}$ and $\delta^2 E_{\lambda^2}$, the linear term in the spin-orbit coupling has the form
 \begin{equation}
     \delta^2 E_{\lambda} = E_{onsite}^{(1)} + E_{intersite}^{(0,1)} + E_{intersite}^{(1,0)} \text{  .}
 \end{equation}
 \noindent
The quadratic term is
 \begin{equation}
     \delta^2 E_{\lambda^2} = E_{onsite}^{(2)} + E_{intersite}^{(0,2)} + E_{intersite}^{(2,0)} + E_{intersite}^{(1,1)} \text{  .}
 \end{equation}
 \noindent
 Note that $\alpha_{ij}^{(0,1)}=\alpha_{ij}^{(1,0)}=\alpha_{ij}^{(0,2)}=\alpha_{ij}^{(2,0)}=\beta_{ij}^{(1,0)}=\beta_{ij}^{(0,1)}=\beta_{ij}^{(0,2)}=\beta_{ij}^{(2,0)}=0$, because $G^{(0)}_{ij}=G_{ij}$ is diagonal in spin basis. Therefore, up to linear order in the spin-orbit coupling, only appears a correction to the isotropic Heisenberg exchange, and possibly an anti-symmetric exchange $d_{ij}$. However, when quadratic terms in spin orbit coupling are included, anomalous terms that breaks the $x-y$ isotropy and mix $x$ and $y$ spin-components can be present.\\
 
 Let' us define $\alpha_{ij} = \alpha_{ij}^{(1,1)}$, and $\beta_{ij}=\beta_{ij}^{(1,1)}$ because these are the only non zero contributions with that structure. Also we define the following linear terms in the spin-orbit coupling:
 \begin{equation}
 \begin{aligned}
 J_{ij}^{(1)}&= J_{ij}^{(0,1)}+ J_{ij}^{(1,0)}\text{  ,}\\
 d_{ij}^{(1)} &= d_{ij}^{(0,1)} +d_{ij}^{(1,0)}\text{  .}
 \end{aligned}
 \end{equation}
 and in analogous way, the quadratic terms in spin orbit coupling can be grouped as
 \begin{equation}
 \begin{aligned}
     J_{ij}^{(2)} &= J_{ij}^{(0,2)}+J_{ij}^{(2,0)}+J_{ij}^{(1,1)} \text{  ,}\\
     d_{ij}^{(2)} &= d_{ij}^{(0,2)}+d_{ij}^{(2,0)}+d_{ij}^{(1,1)}\text{  .}
 \end{aligned}
 \end{equation}

We finally write the isotropic exchange, the anti-symmetric exchange, the on-site constant and canting field as:
 
 \begin{equation}
     \begin{aligned}
         J_{ij} &= J_{ij}^{(0)} + J_{ij}^{(1)} + J_{ij}^{(2)} \text{  ,}\\
         d_{ij} &= d_{ij}^{(1)} + d_{ij}^{(2)} \text{  ,}\\
         K_{i} &= K_{i}^{(1)} + K_{i}^{(2)} \text{  ,}\\
         \mathbf{h}_i &= \mathbf{h}_{i}^{(1)}+\mathbf{h}_{i}^{(2)} \text{  .}
     \end{aligned}
     \label{magCtes}
 \end{equation}
 
 These magnetic constants, are used to parametrize the matrices of eqs. \eqref{DeltaE} as:
\begin{equation}
 \begin{aligned}
     \mathcal{J}_{ij}
     &=\left(\begin{aligned}
         J_{ij} + \alpha_{ij} && \beta_{ij} + d_{ij} \quad\\
\beta_{ij} - d_{ij} \quad&& J_{ij} -\alpha_{ij}
     \end{aligned} \right)\text{   .}
 \end{aligned}
 \label{Jmatrix}
 \end{equation}

In the same way, matrix $\mathcal{A'}$ has dimensions $2\times2$, and it is defined by:

 \begin{equation}
     \mathcal{A'}=-(K^{(2)} + \frac{1}{2}H_W - \frac{1}{2} J^{(2)}_{00})\sigma_0 \text{  ,}
     \label{Ap_green}
 \end{equation}
where $K^{(2)}$ is an onsite term, defined in appendix B, which involves second order processes in the SOC. Moreover, $H_W^i=\sum_{j, (j\neq i)} J_{ij}^{(0)}$ is the Weiss field, that comes from the isotropic part of exchange. In the last expression \eqref{Ap_green}, we used $\sigma_0$ to denote the $2\times 2$ identity matrix. Note that in \eqref{Ap_green} we did not write the site index $i$ explicitly, to make emphasis on that $\mathcal{A'}$ is the same on all magnetic sites.

After the numerical calculation, we test that $d_{ij}$ is zero for all  NN links, consistent with the fact that inversion symmetry is preserved in $CrI_3$. Furthermore, the effective field $\mathbf{h}_i$ was found to be zero on all magnetic sites. This is consistent with the choice of the ground state polarized in $\mathbf{\hat{z}}$. If there had been a nonzero effective field, the  state with magnetization pointing normal to the crystal plane would have been unstable.

 \section{Variation of the generic spin Hamiltonian}
 
 Now, we return to the energy variation of the generic spin Hamiltonian \eqref{Hs}. If we perturb the magnetic ground state, by including tranverse spin components $\delta S_i^x$ and $\delta S_i^y$, and then we expand the Hamiltonian up to quadratic order in these perturbations, we get:
\begin{equation}
\begin{aligned}
    \Delta H_S =& -\sum_{i} \mathbf{h}_i\cdot \delta \mathbf{S}_i - \sum_{i} \delta \mathbf{S}_i \cdot \mathcal{A'}_{ii} \cdot \delta \mathbf{S}_i\\ &-\frac{1}{2} \sum_{i,j} \delta \mathbf{S}_i \cdot \mathcal{J}_{ij}\cdot \delta \mathbf{S}_j\text{ ,}
\end{aligned}
    \label{DeltaHS}
\end{equation}

where $\delta \mathbf{S}_i = (\delta S_i^x, \quad \delta S_i^y)$ is the spin variation. There is a linear term, proportional to a transverse field $\mathbf{h}_i$, whose componetes are defined by:

\begin{equation}
    \mathbf{h}_i = 2S\mathbf{\hat{z}}\cdot \mathcal{A}_{ii} + \sum_{j\neq i}S \mathbf{\hat{z}}\cdot \mathcal{J}_{ij} \text{  .}
    \label{h}
\end{equation}

Also, the onsite $2\times2$ matrix $\mathcal{A'}_{ii}$ is defined by:
\begin{equation}
    \mathcal{A'}_{ii} = \mathcal{A}_{ii} - \left(\mathcal{A}_{ii}^{zz}+\frac{1}{2}\sum_{j\neq i}\mathcal{J}_{ij}^{zz} \right)\sigma_0\text{ .}
    \label{A'}
\end{equation}

In the first term on the right side of eq. \eqref{A'}, only the $x,y$-components of $\mathcal{A}$ are involved. Moreover, we are only interested in the $x$ and $y$-components of $\mathbf{h}_i$ in eq. \eqref{h}. In the same way, on the third term of the right side of eq. \eqref{DeltaHS}, only participates the $x,y$ block of $\mathcal{J}_{ij}$.

The idea is to map the variation \eqref{DeltaE} to the quadratic Hamiltonian \eqref{DeltaHS}. The $2\times2$ exchange matrix $\mathcal{J}_{ij}$ of \eqref{DeltaE} are directly the normal-to-$\mathbf{\hat{z}}$ components of the full exchange matrix in the spin Hamiltonian \eqref{Hs}. Moreover, if we compare eq. \eqref{Ap_green} with the form of the onsite matrix \eqref{A'}, obtained from the spin Hamiltonian, it is easy to note that $\mathcal{A'}$ is a diagonal matrix. As a contribution proportional to the identity on $\mathbf{A}$ in \eqref{Hs} does not depend of the magnetization field, we claim that the only nonzero element of $\mathcal{A}_{ii}$ is $\mathcal{A}_{ii}^{zz}$. Then, in \eqref{Ap_green} could be a contribution from a easy-axis anisotropy, and some contribution that came from the Weiss field (term proportional to the exchange constant, summed over all neighbours of a given site).
\twocolumngrid

\bibliography{CrI3_topology}

\end{document}